\documentclass[amsmath,amssymb,twocolumn,]{revtex4}

\usepackage{graphicx}

\begin{document}

\title{Density of states in SF bilayers with arbitrary strength of magnetic scattering}

\author{D.\,Yu.~Gusakova$^*$,\/\thanks{e-mail: dariamessage@yandex.ru} A.\,A.~Golubov$^+$,
 M.\,Yu.~Kupriyanov$^*$, and
A.~Buzdin$^{\ddag}$}

\address{
$^*$ Nuclear Physics Institute, Moscow State University, 119992 Moscow, Russia\\
$^+$ Department of Applied Physics, University of Twente, 7500 AE, Enschede, The Netherlands\\
$^\ddag$ Institut Universitaire de France and Condensed Matter
Theory Group, CPMOH, University Bordeaux 1, UMR 5798, CNRS, F-33405
Talence Cedex, France
\\}

\begin{abstract}
We developed the self-consistent method  for the
calculation of the density
 of states $N(\varepsilon)$ in the SF bilayers. It based on the quasi-classical
Usadel equations and takes into account the suppression of
superconductivity in
 the S layer due to the proximity effect with the F metal, as well as existing mechanisms of the spin
dependent electron scattering. We demonstrate that the increase of
the spin orbit or spin flip electron scattering  rates results in
completely different transformations of  $N(\varepsilon)$ at the
free F layer interface. The developed  formalism has been applied
for the interpretation of the available experimental data.
\end{abstract}

\maketitle

 It is well known that superconductivity can
be induced into a non-superconducting material from a
superconductor due to proximity effect. In the superconductor (S)
- normal metal (N) bilayers thus induced minigap and the shape of
the density of states (DOS) at free N metal interface  $
N(\varepsilon )$  depends on the values of the suppression
parameters of SN interface and relation between the N layer
thickness and the decay length of the N metal
\cite{GolubovNUM3}-\cite{Belzig}.  The existence of the minigap
has been confirmed experimentally in a variety of the proximity SN
systems (see e.g. \cite{Gueron}-\cite{Cretinon} and references
therein). In the superconductor - ferromagnet (F) bilayers there
are additional bulk F-layer parameters which influence on
$N(\varepsilon )$. They are the exchange field $H$ and the
electron spin scattering processes. The exchange field tends to
align all electron spins along the field axis. It splits the
minigap and density of states for spin up and spin down electrons
(see \cite{Buzdin2005}-\cite{Bergeret2005} for the reviews). The
experimental study of the proximity effect in SF
systems\cite{Kontos}-\cite{Cirillo} and Josephson effect in SFS
junctions \cite{Ryazanov},\cite{Fuare} reveals that besides the
exchange field the additional pair-breaking magnetic mechanism,
namely, spin dependent electron scattering, should be taken into
account for the data interpretation.

There are three types of the spin dependent electron scattering in
the ferromagnet - spin-orbit interaction and spin-flip processes
which may happen along the exchange field direction and in the
plane perpendicular to it. Previously the influence of the
parallel spin-flip and spin-orbit scattering mechanisms on
$N(\varepsilon )$ had been considered in some limiting cases
(rigid boundary conditions at SF interface, limits of large or
small F layer thickness)\cite{Arnold,DeWeert}.

In this paper we for the first time developed the self-consistent
method for the calculation of the density of states in SF
bilayers. It is based on the quasi-classical Usadel equations and
takes into account the suppression of the superconductivity in the
S layer due to the proximity effect with the F metal, as well as
all three mechanisms of the spin dependent electron scattering. We
have demonstrated that the developed formalism can be applied for
understanding the $N(\varepsilon )$ data obtained in
superconductors with the antiferromagnet ordering.
 We consider the SF bilayer consisting of two dirty metals. They are
a superconductor of the thickness $d_s$ and a thin ferromagnet
$d_f$ adjoined at $x=0$. All physical quantities depend on
coordinate $x$ perpendicular to the SF boundary. The exchange
field is parallel to the SF interface plane. DOS can be calculated
from the Usadel equations. To proceed further it is convenient to
use the $\theta$ parametrization
$G(\omega,x)=\cos\theta(\omega,x)$,
$F(\omega,x)=\sin\theta(\omega,x)$, where $G$ and $F$ are normal
and anomalous Green functions. The magnetic and spin-orbit
scattering mix up the up and down spin states which couples the
Usadel equations for the Green functions with the opposite spin
directions. In the F layer ($x<0$) it gives the system of the two
equations
\begin{multline}\label{01}
-\frac{D_f}{2}\frac{\partial^2\theta_{f1(2)}}{\partial x^2}+\left(
\omega \pm i
H+\frac{1}{\tau_z}\cos\theta_{f1(2)}\right)\sin\theta_{f1(2)}+\\
+\frac{1}{\tau_x}\sin(\theta_{f1}+\theta_{f2}) \pm
\frac{1}{\tau_{so}}\sin(\theta_{f1}-\theta_{f2})=0,
\end{multline}
%\begin{multline}\label{01}
%-\frac{D_f}{2}\frac{\partial^2\theta_{f2}}{\partial x^2}+\left(
%\omega - i
%H+\frac{1}{\tau_z}\cos\theta_{f2}\right)\sin\theta_{f2}+\\
%+\frac{1}{\tau_x}\sin(\theta_{f1}+\theta_{f2}) -
%\frac{1}{\tau_{so}}\sin(\theta_{f1}-\theta_{f2})=0,
%\end{multline}
and in the S layer ($x>0$) the Usadel equations stay uncoupled

\begin{equation}\label{02}
-\frac{D_s}{2}\frac{\partial^2\theta_{s1(2)}}{\partial x^2}+
\omega \sin\theta_{s1(2)}=\Delta(x) \cos\theta_{s1(2)},
\end{equation}
where $\theta_1$ and $\theta_2$ correspond to the Green functions
with the opposite spin directions, $\omega=\pi T(2n+1)$ are the
Matsubara frequencies, $D_s (D_f)$ is the diffusion coefficient in
S (F) layer, $H$ is the exchange field energy in F layer,
$\Delta(x)$ is the superconducting energy gap which is zero in F
layer. Here we use the self-consistent method to resolve the
Usadel equations which takes into account the decrease of the
energy gap $\Delta$ in the S layer from its bulk value along
$x$-axis towards the boundary due to the proximity effect. The
scattering times are labelled here as $\tau_z$, $\tau_x$ and
$\tau_{so}$, where $\tau_{z(x)}$ corresponds to the magnetic
scattering parallel (perpendicular) to the quantization axis and
$\tau_{so}$ corresponds to the spin-orbit scattering.

In the S layer the Usadel equations are completed with the
self-consistency equation
\begin{equation}\label{03}
\Delta(x)\ln t+t\sum_{\omega=0}^{\omega=\infty}\left[
\frac{2\Delta(x)}{\omega}-\sin\theta_{s1}-\sin\theta_{s2}\right]=0,
\end{equation}
where $t=T/T_c$, $T_c$ is the bulk superconducting temperature.
Here and further we work with the normalized energy parameters
$\Delta\equiv\Delta/\pi T_c$, $\omega\equiv\omega/\pi T_c$,
$H\equiv H/\pi T_c$, and for the length parameters in F layer
$x\equiv x/\xi_n$, $\xi_n=\sqrt{D_f/2\pi T_c}$, and in S layer
$x\equiv x/\xi_s$, $\xi_s=\sqrt{D_s/2\pi T_c}$. The scattering
parameter notations are $\alpha_z=(\tau_z \pi T_c)^{-1}$,
$\alpha_x=(\tau_x \pi T_c)^{-1}$, $\alpha_{so}=(\tau_{so} \pi
T_c)^{-1}$.

The boundary conditions at FS interface have the form
\begin{equation*}
\gamma_B  \frac{\partial \theta_{f1(2)}}{\partial x} \mid_{x=-0}
=\sin (\theta_{s1(2)}-\theta_{f1(2)}),
\end{equation*}
\begin{equation}\label{04}
\frac{\gamma_B}{\gamma}  \frac{\partial \theta_{s1(2)}}{\partial
x} \mid_{x=+0} =\sin (\theta_{s1(2)}-\theta_{f1(2)}),
\end{equation}
and at free edges
\begin{equation}\label{05}
\frac{\partial\theta_{f1(2)}}{\partial
x}\mid_{x=-d_f}=0,\,\frac{\partial\theta_{s1(2)}}{\partial
x}\mid_{x=d_s}=0,
\end{equation}
%\begin{equation}\label{05}
%\frac{\partial\theta_{s1(2)}}{\partial x}\mid_{x=d_s}=0,
%\end{equation}
where $\gamma=(\sigma_n \xi_s)/(\sigma_s \xi_n)$, $\sigma_{n(s)}$
is the conductivity of the F(S) layer, $\gamma_B=\frac{R_b
\sigma_n}{\xi_n}$, $R_b$ is the specific resistance of the SF
interface.

\begin{figure}[h]
 \centerline{\includegraphics[width=0.5\textwidth]{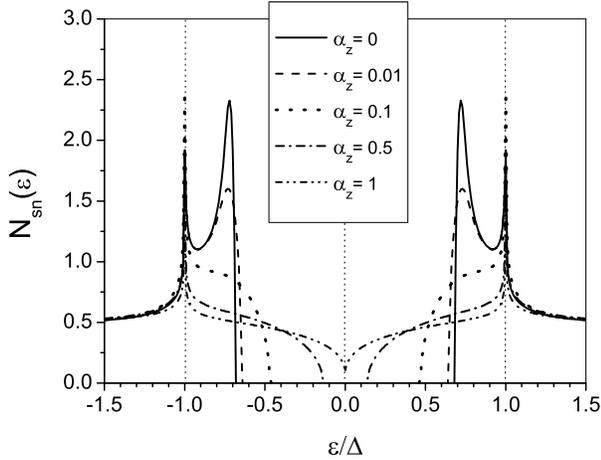}}
 \caption{The case of the parallel magnetic scattering in the S/N bilayer ($H/\pi T_c=0$).
 Spin up energy DOS variation in the N layer
 for  several values of the magnetic scattering parameter $\alpha_z$
 and for the fixed $\gamma_B=5$, $\gamma=0.05$, $d_f/\xi_n=0.2$,
 $d_s/\xi_s=10$, $\alpha_x=0$, $\alpha_{so}=0$.}
 \label{fig:fig2}
\end{figure}
 For the arbitrary layers
thicknesses, interface parameters, $\gamma$, $\gamma_B$, and
magnetic scattering parameters the equations (\ref{01})-(\ref{05})
have been solved numerically using the self-consistent two step
iterative procedure (for ref. see
\cite{GolubovNUM3,GolubovNUM1,GolubovNUM2}). In the first step we
calculate the order parameter coordinate dependence
 $\Delta(x)$, in  the Matsubara technique using the self-consistent condition in the
S layer. Due to the proximity effect $\Delta(x)$  decreases
towards the SF interface. Then by proceeding to the analytical
extension in (\ref{01}), (\ref{02}) over the energy parameter
$\omega\rightarrow -i\varepsilon$ and using $\Delta(x)$ dependence
obtained in the previous step we find the Green functions by
repeating the iterations until the convergency is reached. The
density of states
$N(\varepsilon)=N_\uparrow(\varepsilon)+N_\downarrow(\varepsilon)$
can be found as
\begin{equation}\label{06}
N_{\uparrow(\downarrow)}(\varepsilon)=0.5 N(0)Re
\cos\theta_{1(2)},
\end{equation}
%\begin{equation}
%N(\varepsilon)=N_\uparrow(\varepsilon)+N_\downarrow(\varepsilon),
%\end{equation}
where $N_{\uparrow(\downarrow)}$ is the DOS for the one spin
direction and $N$ is the total DOS.

\begin{figure}[h]
 \centerline{\includegraphics[width=0.5\textwidth]{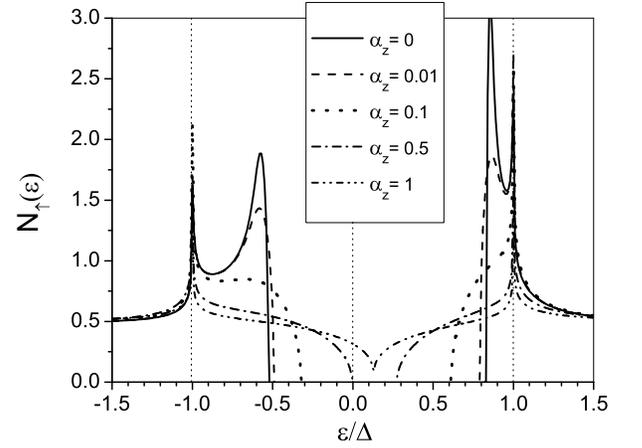}}
 \caption{The case of parallel magnetic scattering. Spin up Energy DOS variation
 in the F layer
  for several values of the magnetic scattering parameter $\alpha_z$
 and for the fixed $H/\pi T_c=0.2$, $\gamma_B=5$, $\gamma=0.05$, $d_f/\xi_n=0.2$,
 $d_s/\xi_s=10$, $\alpha_x=0$, $\alpha_{so}=0$.}
 \label{fig:fig3}
\end{figure}

\begin{figure}[h]
 \centerline{\includegraphics[width=0.5\textwidth]{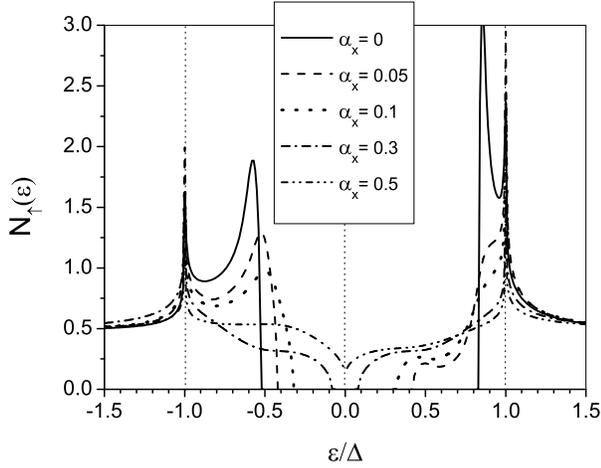}}
 \caption{The case of the perpendicular magnetic scattering.
 Spin up energy DOS variation in the F layer
  for several values of the magnetic scattering parameter $\alpha_x$
 and for the fixed $H/\pi T_c=0.2$, $\gamma_B=5$, $\gamma=0.05$, $d_f/\xi_n=0.2$,
 $d_s/\xi_s=10$, $\alpha_z=0$, $\alpha_{so}$=0.}
 \label{fig:fig4}
\end{figure}

\begin{figure}[h]
 \centerline{\includegraphics[width=0.5\textwidth]{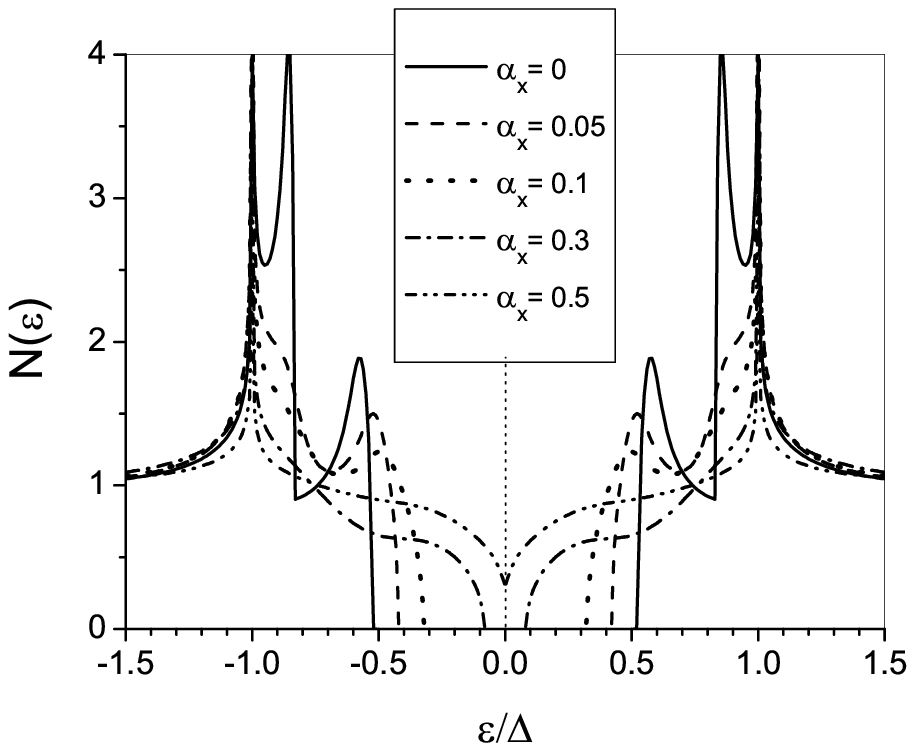}}
 \caption{The case of the perpendicular magnetic scattering.
 Total DOS energy variation
 in the F layer  for  several values of the magnetic scattering parameter $\alpha_x$
 and for the fixed $H/\pi T_c=0.2$, $\gamma_B=5$, $\gamma=0.05$, $d_f/\xi_n=0.2$,
 $d_s/\xi_s=10$, $\alpha_z=0$, $\alpha_{so}$=0.}
 \label{fig:fig5}
\end{figure}

\begin{figure}[h]
 \centerline{\includegraphics[width=0.5\textwidth]{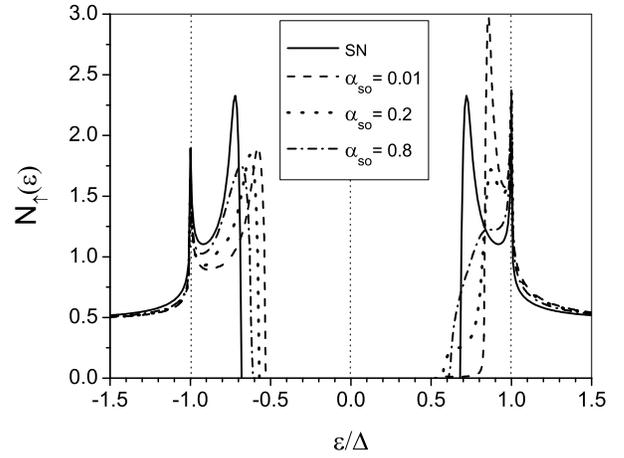}}
 \caption{The case of spin-orbit scattering.  Spin up energy DOS variation
 in the F layer
  for  several values of the magnetic scattering parameter $\alpha_{so}$
 and for the fixed $H/\pi T_c=0.2$, $\gamma_B=5$, $\gamma=0.05$, $d_f/\xi_n=0.2$,
 $d_s/\xi_s=10$, $\alpha_z=0$, $\alpha_{x}$=0. Black curve
 corresponds to the SN case.}
 \label{fig:fig6}
\end{figure}

\begin{figure}[h]
 \centerline{\includegraphics[width=0.5\textwidth]{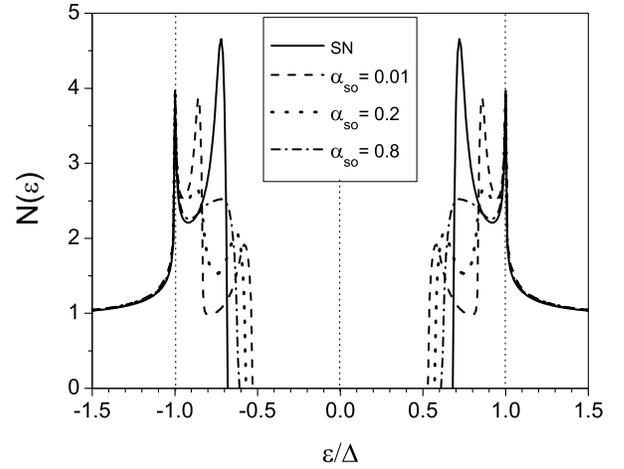}}
 \caption{The case of spin-orbit scattering. Total DOS energy variation
  in the F layer
  for several values of the magnetic scattering parameter $\alpha_{so}$
 and for the fixed $H/\pi T_c=0.2$, $\gamma_B=5$, $\gamma=0.05$, $d_f/\xi_n=0.2$,
 $d_s/\xi_s=10$, $\alpha_z=0$, $\alpha_{x}$=0. Black curve
 corresponds to the SN case.
}
 \label{fig:fig7}
\end{figure}

\begin{figure}[h]
 \centerline{\includegraphics[width=0.5\textwidth]{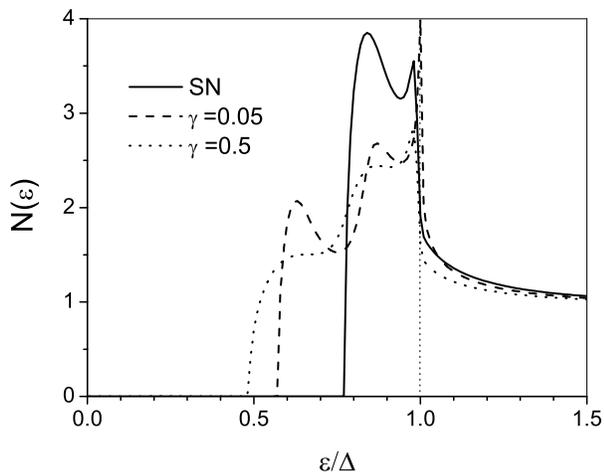}}
 \caption{The case of spin-orbit scattering. Total DOS energy variation
 in the F layer
  for $\tau_{so}^{-1}=H$ and for the fixed $H/\pi T_c=0.2$, $\gamma_B=5$, $d_f/\xi_n=0.2$,
 $d_s/\xi_s=10$, $\alpha_z=0$, $\alpha_{x}$=0. Black curve
 corresponds to the SN case with $\gamma=0.5$. Dashed and dotted
 curves correspond to two different values of the
 $\gamma$ parameter.}
 \label{fig:fig8}
\end{figure}

The numerically obtained energy dependencies of DOS in F layer at
the free F boundary are presented on Fig.
\ref{fig:fig2}-\ref{fig:fig8}. At $H=0$ (fig. \ref{fig:fig2}) we
reproduce the well known mini gap existing in SN bilayer
\cite{GolubovNUM3,GolubovNUM1,GolubovNUM2}. The presence of the
unixial magnetic scattering tends to smooth the BCS peaks in the
DOS. Figure \ref{fig:fig3} demonstrates the DOS evolution for the
spin up electrons for different parameters $\alpha_z$ where the
full black curve corresponds to the usual splitted peaks within
the energy gap due to the exchange field in the absence of any
magnetic scattering. By adding the magnetic scattering aligned
with the exchange field direction one can see the smearing of the
sharp peaks with the gradual closing of the induced energy gap in
the F layer. It is interesting to note that the symmetry of the
spin resolved DOS in respect of Fermi energy ($\varepsilon=0$)
does not exist in the presence of magnetic scattering.

Figure \ref{fig:fig4} demonstrates the influence of the
perpendicular magnetic scattering on the energy DOS variation
within the energy gap. Total DOS for both spin directions for the
different values of the perpendicular magnetic scattering
parameter $\alpha_x$ is plotted in Fig. \ref{fig:fig5}. The peaks
in DOS are slowly moving towards the zero energy that can be
explained as the presence of some additional splitting field
besides the ordinary exchange field in ferromagnet. As in the case
of parallel magnetic scattering the peaks are smoothed out and the
energy gap disappears. For the small magnetic scattering times
$\tau_z$ and $\tau_x$ the DOS tends to its bulk value in the
ferromagnet.

Figures \ref{fig:fig6} and \ref{fig:fig7} depict the spin up and
total DOS for different parameters of spin-orbit scattering,
correspondingly. It can be seen that in contrast to the magnetic
scattering described above the spin-orbit scattering tends to
decrease the effect of the peak splitting within the energy gap
cased by the ferromagnetic exchange field. Black curves in Fig.
\ref{fig:fig6} and \ref{fig:fig7} correspond to the zero exchange
field (SN structure case). The smaller the spin-orbit scattering
time the closer the curve to the superconductor/normal metal case
and two minigap behavior degenerate to the one minigap curve as in
the SN structure.

It is interesting to mention the peculiarity in DOS dependence in
the presence of the spin-orbit scattering. As it was shown in
\cite{Demler1997,Kuplev1990} in the presence of the spin-orbit
scattering for the parameter $\tau_{so}^{-1}=H$ the solution of
the Usadel equation changes its characteristic behavior from the
oscillating one to the damping decay, that should also cause the
changes in the energy DOS variation. Figure \ref{fig:fig8}
demonstrates the appearance of the plateau instead  of peak in DOS
for $\tau_{so}^{-1}=H$ for some parameter $\gamma$ when it is
large enough to diminish \, the penetration of superconductivity
into the F layer. For the particular set of parameters ($H,
\gamma_B, d_s, d_f$) used for calculation of the graph in Fig.
\ref{fig:fig8} this transformation occurs approximately at
$\gamma\approx0.5$.

Recently,  the coexistence of the magnetic and superconducting
order in nickel borocarbides was studied in several laboratories
experimentally. Such compounds as $ErNi_2B_2C$ and $TmNi_2B_2C$
both being the superconducting materials demonstrate radically
different magnetic properties. Local tunnelling microscopy at low
temperatures revealed considerable difference in the local
superconducting density of states behavior. In contrast with
$TmNi_2B_2C$ \cite{Suderow} compound where DOS has its usual BCS
type, $ErNi_2B_2C$ \cite{Crespo} measurements show the non zero
conductance and thereby the non zero DOS within the energy gap.

To find the possible explanation of such a difference we propose
the following model. We believe that in $ErNi_2B_2C$ compound the
magnetic order near the surface is absent even when the
antiferromagnetic phase appears in the bulk. This may be related
with some atomic compositional disorder near the surface and
modified exchange interaction between magnetic moments near the
surface. Consequently, to describe the surface properties of
superconducting $ErNi_2B_2C$, the model of a thin film with the
relatively strong magnetic scattering on the top of the bulk
superconductor without magnetic scattering seems to be quite
reasonable.

\begin{figure}[h]
 \centerline{\includegraphics[width=0.4\textwidth]{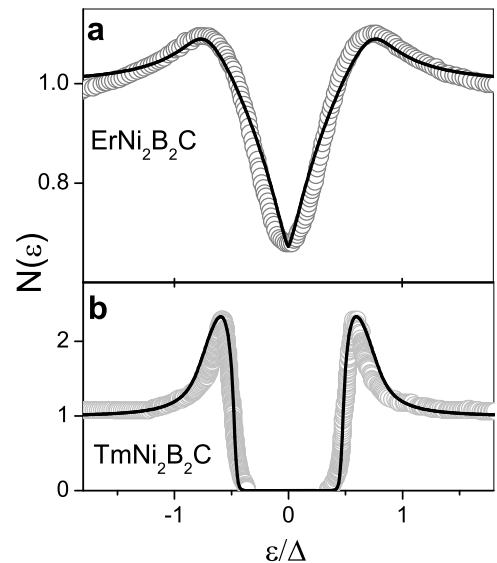}}
 \caption{Theoretical fit of the experimental data of \cite{Crespo}(Fig.1).
 Plot parameters: $H=0$, $\gamma=1$, $\gamma_B=0$, $d_s\gg\xi_s$, $\alpha_z=\alpha_{so}=0$.
 \textbf{a}) $T=0.15K$,  $T_c=11K$, $d_f=0.35\xi_n$
 and magnetic scattering parameter $\alpha_x=0.95$;
 \textbf{b}) $T=0.8K$, $T_c=10.5K$, $d_f=0.6\xi_n$,
 no magnetic scattering. }
 \label{fig:fig9}
\end{figure}

Using the  developed algorithm for the SF bilayer, we may assume
the exchange field $H=0$ as in the paramagnetic case and
$\gamma=1$, $\gamma_B=0$ for the actual absence of the boundary.
Fig. \ref{fig:fig9}a demonstrates the calculated DOS behavior at
$x=-d_f$ in the presence of the magnetic scattering which destroys
the usual BCS behavior. For  $ErNi_2B_2C$ having easy plane
magnetic anisotropy we take $1/\tau_z=0$ and
$1/\tau_x=1/\tau_y=1/\tau$.
%Note that in case of the zero exchange
%field $H=0$ there is no difference between the parallel and
%perpendicular magnetic scattering, so we have the case of the
%isotropic magnetic scattering $1/\tau_z=1/\tau_x=1/\tau$.
Fig.\ref{fig:fig9}b corresponds to the case without magnetic
scattering. It can be seen that both black theoretical curves are
in a good agreement with the experimental data of \cite{Crespo}
(Fig.1a and Fig.2b). The difference between $ErNi_2B_2C$ an
$TmNi_2B_2C$ curves may be related with the important difference
in their Neel temperatures ($6$ K and $1.5$ K, respectively). The
lower $T_N$ may lead to the much smaller magnetic scattering in
$TmNi_2B_2C$.

In conclusion we demonstrate that the increase of spin orbit or
spin flip electron scattering  rates results in completely
different transformations of  $N(\varepsilon)$ at free F layer
interface. The increase of $\tau_z^{-1}$ results in the continuous
suppression of the peaks in the density of states accompanied by
the closing of the energy gap. The increase of $\tau_x^{-1}$
additionally leads to the shift of the peaks towards the zero
energy which looks like the action of some additional exchange
field in the ferromagnet. Contrary to that the increase of
$\tau_{so}^{-1}$ does not result in the closing of the energy gap
and tends to decrease the Zeeman peaks splitting.

All calculations have been done in a self-consistent way in the
frame of the Usadel equations.
 The developed
formalism has been successfully applied for the interpretation of
the data obtained in the superconductors with the antiferromagnet
ordering.

This work has been supported by RFBR project 06-02-90865 We
acknowledge the support by French EGIDE programme 10197RC, ESF
PI-Shift Programme and NanoNed programme under project TCS 7029.
We are grateful to H. Suderow for useful discussions and providing
experimental data prior to publication.

\end{document}